\documentstyle[12pt]{article}

\newcommand{\bmat}{\left(\begin{array}}
\newcommand{\emat}{\end{array}\right)}

\def\yzero{\smash{\hbox{$y\kern-4pt\raise1pt\hbox{${}^\circ$}$}}}

\def\beq{\begin{equation}}
\def\eeq{\end{equation}}
\def\beqa{\begin{eqnarray}}
\def\eeqa{\end{eqnarray}}

\def\-{\hphantom{-}}

\def\s2{\frac{1}{\sqrt2}}

\def\beq{\begin{equation}}
\def\eeq{\end{equation}}
\def\beqa{\begin{eqnarray}}
\def\eeqa{\end{eqnarray}}

\def\IF{\relax{\rm I\kern-.18em F}}
\def\II{\relax{\rm I\kern-.18em I}}
\def\IP{\relax{\rm I\kern-.18em P}}
\def\IC{\relax\hbox{\kern.25em$\inbar\kern-.3em{\rm C}$}}
\def\IR{\relax{\rm I\kern-.18em R}}

\def\cp{{\cal P}}

\def\Dsl{\,\raise.15ex\hbox{/}\mkern-13.5mu D} 
\def\IZ{Z\kern-.4em  Z}

 \def\cp#1{\relax\ifmmode {\IP\kern-2pt{}_{#1}}\else $\IP\kern-2pt{}_{#1}$\=fi}

\catcode`\@=11
\newdimen\@rotdimen
\newbox\@rotbox

\def\@vspec#1{\special{ps:#1}}
\def\@rotstart#1{\@vspec{gsave currentpoint currentpoint translate
   #1 neg exch neg exch translate}}
\def\@rotfinish{\@vspec{currentpoint grestore moveto}}
\def\@rotr#1{\@rotdimen=\ht#1\advance\@rotdimen by\dp#1%
   \hbox to\@rotdimen{\hskip\ht#1\vbox to\wd#1{\@rotstart{90 rotate}%
   \box#1\vss}\hss}\@rotfinish}
\def\@rotl#1{\@rotdimen=\ht#1\advance\@rotdimen by\dp#1%
   \hbox to\@rotdimen{\vbox to\wd#1{\vskip\wd#1\@rotstart{270 rotate}%
   \box#1\vss}\hss}\@rotfinish}%
\def\@rotu#1{\@rotdimen=\ht#1\advance\@rotdimen by\dp#1%
   \hbox to\wd#1{\hskip\wd#1\vbox to\@rotdimen{\vskip\@rotdimen
   \@rotstart{-1 dup scale}\box#1\vss}\hss}\@rotfinish}%
\def\@rotf#1{\hbox to\wd#1{\hskip\wd#1\@rotstart{-1 1 scale}%
   \box#1\hss}\@rotfinish}%
\def\rotate{\@ifnextchar[{\@rotate}{\@rotate[l]}}
\def\@rotate[#1]#2{\setbox\@rotbox=\hbox{#2}\@nameuse{@rot#1}\@rotbox}

\catcode`\@=12

\begin{document}

\makeatletter \@addtoreset{equation}{section} \makeatother
\renewcommand{\theequation}{\thesection.\arabic{equation}}
\pagestyle{empty}
\rightline{\tt CK-TH-2000-003}
\rightline{\tt hep-th/0006194}
\vspace{0.5cm}
\begin{center}
\LARGE{\bf   
Constraint Supersymmetry Breaking and Non-Perturbative
Effects in String Theory}\\[10mm]
\medskip
\large{C.~Kokorelis$^{1}$
\\[2mm]}
\small{$^1$ 
Center for Mathematical Trading and Finance, CITY University,\\[-0.3em]
Frobischer Crescent, Barbican Centre,
London, EC2Y 8HB, U.K.
\\[6mm]}
\small{\bf Abstract} \\[7mm]
\end{center}

\begin{center}
\begin{minipage}[h]{14.5cm}
We discuss supersymmetry breaking mechanisms
at the level of low energy ${\cal N }= 1$ effective heterotic superstring actions
that exhibit
$SL(2,Z)_T$ target space modular duality or
$SL(2,Z)_S$
strong-weak coupling duality.
The allowed superpotential forms
use the assumption that the sourse of non-perturbative effects is
not specified and as a result represent the most
general parametrization of non-perturbative effects.
We found that the allowed non-perturbative superpotential
is severely constrained when we use the cusp forms of
the modular group for its construction. By construction the poles of the
superpotential are either inside the fundamental domain or beyond.
We also found limits on the parameters of the superpotential
by demanding that the
truncated potential for the gaugino condensate never breaks down
at finite values in the moduli space.
The latter constitutes a criterion for avoiding poles in the fundamental
domain. However, the potential in most of the cases avoids  
naturally singularities inside the
fundamental domain, rendering the potential finite.
The minimum values of the limits on the parameters in the superpotential may
correspond
to vacua with vanishing cosmological constant.
\end{minipage}
\end{center}
\newpage
\setcounter{page}{1} \pagestyle{plain}
\renewcommand{\thefootnote}{\arabic{footnote}}
\setcounter{footnote}{0}

\newpage

\section{\bf Introduction}

One of the biggest problems that heterotic string theory, and its "equivalents", e.g type II,
I,
have to face today
 is the question
of ${\cal  N} =1$ space-time supersymmetry breaking. The breaking, due to the
presence of the gravitino, that determines the scale of supersymmetry
breaking,
in the effective action, must be
spontaneous and not explicit. 
Several mechanisms have been used in
recent years to break consistently supersymmetry.
They can distinguished as to when
they are at work at the string theory level or at the
effective superstring action level.
The first category of mechanisms includes the tree level coordinate dependent
compactification mechanism \cite{CDC}, the
magnetized tori approach \cite{bachas}, the type I brane
breaking \cite{apo1, angel, apo2}, the partial breaking \cite{tayloo}
while the latter category includes approaches that use target space
duality e.g \cite{feto, iba1, bigai, cve} or S-duality \cite{iba2, homou} at the
level of
effective superstring action, related to gaugino condensation \cite{vene},
to constrain
 the allowed superpotential forms.
The main problem in all approaches is the creation of 
an appropriate potential for the moduli and the dilaton that 
can fix their vacuum expectation values.
The first category was made popular quite recently because of our
understanding of non-perturbative effects in string theory via the discovery of
D-branes, oblects where open strings can end.
However, we might not forget that gaugino condensation, a non-perturbative field
theoretical effect can break supersymmetry, satisfactory espacially when two
condensates are used \cite{de}. In fact breaking supersymmetry
by gaugino condensation (GC) has its advantages. In fact if we knew
the non-perturbative contribution to the gauge kinetic function, GC
would have been an excellent mechanism and we wouldn't have to look
on string mechanisms
to break supersymmetry. Here, we would do exactly that. Because
heterotic string theory has target space duality as one of its properties
we can use modular forms to parametrize the unknown non-perturbative
dynamics \cite{cve}, practically to parametrize the unknown 
non-perturbative contributions to the gauge kinetic function f.
Something similar could not be done e.g for type I
strings as they don't possess target space duality 
so a string breaking of supersymmetry for the latter
may be the most appropriate.

The purpose of this paper is to reexamine the
issue of constructing superpotentials $W^{np}$ that affect supersymmetry
breaking at the level of ${\cal N} = 1$ effective heterotic superstring actions
when the sourse of non-perturbative effects, is not specified.
We will not perform a full 
a numerical study of the scenarios proposed as we leave it for future work. 
In particular in this work, we examine and improve in earlier scenaria \cite{iba2, cve, homou}
the assumptions used in the construction of $W^{np}$.

 The modification
of the ${\cal N} = 1$ heterotic effective action that we examine in this work amounts
to modifying the superpotential when T-duality or S-duality
non-perturbative effects are included.
In this work we want to break supersymmetry dynamically, rather
than geometrically, thus we make use of
gaugino condensation in supergravity \cite{tay}.
In general there are two different approaches in describing
gaugino condensation. These are the effective
lagrangian approach \cite{feto, bigai} where we can use a
gauge singlet bilinear superfield U as a dynamical degree of freedom
and the effective superpotential approach \cite{iba1, cve}.
In the latter formalism
the gauge singlet bilinear superfield is integrated out through
its equation of motion. 

The paper is organized as follows.
In section 2 we review the current status of
the most general parametrization of non-perturbative effects into the
vacuum structure of ${\cal N} = 1$ heterotic superstrings through
modifications of the
superpotential in the effective superpotential approach.
In section 3 we describe new parametrizations of non-perturbative effects
 by constructing the most general weight zero superpotential
factor invariant under $SL(2,Z)$ modular transformations.
It is found, by using cusp modular forms that the resulting superpotential
automatically includes weak coupling limit constraints in its form, when 
it is used to describe S-dual superpotentials.

Moreover we connect previously unrelated constructions with vanishing
cosmological constant and 
broken supersymmetry in the dilaton sector,
auxiliary dilaton field $h_s$ non-zero \cite{cve},
to our non-perturbative superpotential constructions through
basis for modular forms. 

In section 4 we decsribe the most general modifications to
the superpotential
in the effective lagrangian approach of \cite{fmtve} when T-duality
non-perturbative effects are included with
 \cite{luta, lumu} and without
the formation of matter field condensates \cite{fmtve, tay}.
In section 4.3
we test the stability of the condensate dynamics for the
new superpotential constructions.
We found strong constraints on the parameters of the superpotential
such that the truncated approximation never breaks down as the potential
approaches its self-dual points.
We found that the lower
classes
of superpotentials with finite potential at finite points in the
upper half plane may coincide with the vacua with vanishing cosmological constant
mentioned in section 3.
Finally, in section 5 we present our conclusions and some future
directions of this work.

\section{\bf Allowed forms of non-perturbative superpotentials}

The effective supergravity
theory coming from superstrings is described by the knowledge
of three functions, the K\"ahler potential, the superpotential W
and the gauge kinetic function f, that all depend on the moduli fields. 
For $(2,2)$ heterotic string compactifications
with ${\cal N} = 1$ supersymmetry
there is at least
one complex modulus which we denote by $T= R^2 + i b$, where
R is the breathing mode of the six dimensional internal space and b is the internal
axion $\theta$. The T-field corresponds, when T large, to the globally
defined $(1, 1)$ K\"ahler form.  
Here we will restrict our study to the simplest (2, 2) models where there is
a single overall modulus, by freezing all other T-moduli.
In the gravitational sector the lowest component
of the chiral dilaton superfield S forms a complex scalar modulus,
combining the gauge coupling constant as its
real part with the pseudoscalar axion field,
\begin{equation}
S = \frac{1}{g^2} + i \theta.
\label{dila}
\end{equation}
The part of the K\"ahler potential the includes the tree level
contribution\footnote{For simplicity we neglect the Green-Schwarz term
contribution as it does not introduce any additional dilaton dependence
on the K\"ahler potential \cite{fekouluzwi}.
Since there is no known method to calculate the
non-perturbative corrections to the K\"ahler potential, 
we consider the Kahler potential as receiving its tree 
level value.   
}
to the dilaton is 
\begin{equation}
K(S, {\bar S}) = - \log(S + {\bar S}).
\label{asdas1}
\end{equation}
The K\"ahler potential, for the T-field, is defined as
\begin{equation}
K(T, {\bar T})=  -\log( T+ {\bar T})^3
\label{asda01}
\end{equation}
in its tree level form and
\begin{equation}
K(T, {\bar T})=  -\log\{( T+ {\bar T})^3 + {\cal I}_{instanton} \}
\label{asda1}
\end{equation}
in the presence of instantons.

The first term in (\ref{asda1}) dominates in the large radius limit, in
the $\sigma$-model sense, and can be
derived from the field theoretical truncation limit of 10D ${\cal N} = 1$ heterotic
string. The second term represents the contribution of the  
non-perturbative effects and is associated with instantons.

Because moduli fields have flat
potential to all orders of perturbation theory \cite{dix}
their vacuum expectation value's (vev's) remain undetermined. As a result
the task of lifting their degenerate vev's is attributed to
non-perturbative effects. In the absence of a mechanism of calculating
non-perturbative corrections to the K\"ahler potential we can choose to 
include a general parametrization of T-duality effects in the
non-perturbative superpotential that includes the dilaton.
Furthermore we assume that the T-modulus and the dilaton
dependence in the superpotential factorize.
One further constraint in the form of allowed superpotentials comes from its  modular
weight
and the presence of physical singularities in the upper half plane.
All the previous information
can be used to construct non-trivial modular superpotentials with T
or S moduli, that can give
information about the general dynamics of superstring vacua. 
The origin of the perturbative terms in the superpotential
may be understood in the context
of the orbifold limit of 
${\cal N} = 1$ four dimensional  F-theory compactifications on a four-fold
$CY_4$($CY_4 = (CY_3 \times T^2)/Z_2$) dual to $(0,2)$ heterotic
string compactifications on a three-fold Z over a base $F_k$($Z = (K_3
\times T^2)/Z_2$, with a choise of gauge bundle on $E_8 \times E_8$.
In this case, assuming that no heterotic 5-branes,
the F-theory superpotential \cite{wittet}, with no 3-branes present, can match the perturbative
heterotic superpotential \cite{curio}. 
The non-perturbative contributions originate from the 
introduction of type IIB three branes in F-theory necessasy for 
cancelling tadpole matching of
heterotic five-branes(equivalently
we may consider four dimensional M-theory compactifications \cite{dona} on Z $\times S^1/Z_2$,
namely compactifications on a Calabi-Yau threefold Z with
 vector bundle Z embedded in $E_8 \times E_8$.). The latter objects are such that
their four dimensional part
spans the  
four dimensional Minkowski space, while the two remaining
wrap around a holomorphic curve in Z. 
However, in the non-perturbative case 
the question still remains how we can calculate
explicit non-perturbative superpotentials
that translating 
 them in heterotic language we can derive
general conclusions about the potential vacuum structure of heterotic strings.

Because the effective supergravity action of the heterotic string
is invariant
 under
the target space duality transformations, that holds in all orders
of perturbation theory \cite{alva}
\begin{equation}
T \rightarrow \frac{A T - iB}{i C T + D},\; AD - BC = 1,
\label{target}
\end{equation}
since the $G = K + \log|W|^2$ function has to remain invariant, we obtain
that the superpotential has to transform with modular weight -3 \cite{shape}.
Note that if the S-duality principle \cite{iba2, duff, rey, sen} is proved to be
valid principle
at the level of ${\cal N} = 1$ heterotic effective actions the effective
superstring action may be invariant under the $SL(2,S)_S$ transformations
\begin{equation}
S \rightarrow \frac{A^{\prime} S - i B^{\prime}}{i C^{\prime} S + D^{\prime}},\;A^{\prime} D^{\prime} - B^{\prime} C^{\prime}=1,
\label{dual3}
\end{equation}
and the superpotential has to transform with modular weight -1
under (\ref{dual3}).

In \cite{cve}, based on a mathematical theorem of modular forms \cite{lehner},
the most general holomorphic modular
function\footnote{which has no singularities in the fundamental domain}
 of weight r, for a moduli $\Phi$, was written in the form
\begin{equation}
 [G_6(\Phi)]^m [G_4(\Phi)]^n [\eta(\Phi)]^{2r-12m-8n} {\cal P}(j(\Phi)),
\label{megas0}
\end{equation}
or equivalently, the superpotential W,
\begin{eqnarray}
W(\Phi) =  (j-1728)^{m/2} j^{n/3} [\eta(\Phi)]^{2r} {\cal P}(j(\Phi)),\\
W(\Phi) = \Omega(\Phi)[\eta(\Phi)]^{2r} {\cal P}(j(\Phi)),\;\\
\Omega(\Phi)=(j-1728)^{m/2} j^{n/3},
\label{megas00}
\end{eqnarray}
where m, n positive integers and $G_6$ , $G_4$ are the Eisenstein
functions of modular weight six
and four and ${\cal P}(j(\Phi))$ an arbitrary polynomial of the
absolutely modular invariant $j(\Phi)$. 
Depending on whether the superpotential has
modular weight -3 as it is the case of a T-duality invariant
superpotential or modular weight -1 as it is the case of an S-duality
invariant superpotential the classes of superpotentials in (\ref{megas0}-\ref{megas00})
were written in the following forms when r equals -3 \cite{cve}
\begin{equation}
W(T, S) = \frac{(j(T)-1728)^{m/2} j^{n/3}(T) }{\eta^6(T)}{\cal P}(j(T)){\cal K}(S),
\label{duale1}
\end{equation}
where ${\cal K}(S)$ parametrizes the dilaton dynamics,
or -1 \cite{iba2} respectively,
\begin{equation}
W(S) =   \frac{(j(S)-1728)^{m/2} j^{n/3}(S)  }{\eta^2(S)}{\cal P}(j(S)).
\label{duale2}
\end{equation}
\newline
In the last equation we have chosen not to exhibit its T-dependence.
The behaviour of the superpotentials in (\ref{duale1}), 
(\ref{duale2}), is such that the
potential diverges when $T, S \rightarrow \infty$, respectively. However, this
runaway behaviour is avoided as duality stabilizes the potential
at finite points.

In general duality stabilizes the potentials with local minima at 
the points $T, S = 1, \rho$. 
The minima on the cases considered in \cite{iba2, cve} are either at
the self-dual points giving unbroken space-time supersymmetry in the T, S
field sector respectively or in the general case supersymmetry breaking minima 
with negative cosmological constant.                 
In the latter case the minima occur at the boundary of the moduli space.

It is worth mentioning at this point that (\ref{duale1})
is equivalent \cite{cve} to defining\footnote{in the following section $\Omega(T)$
will be replaced by the more general form $\Sigma(T)$}the gauge kinetic function in the form
\begin{eqnarray}
f = S -  \frac{|G_i|}{|G|}(b_a^{N=2}\log[\eta^4(T)(T + {\bar T})]- 
\frac{1}{16\pi^2} Re\{ \partial_T \partial_U h^{(1)}(T, U) - &
\nonumber\\2\log((j(T) -j(U))\}) 
 + (b_a /3) \log|\Omega(T)|^2 + {\cal O}(e^{-S}),
 \label{gauge1}
\end{eqnarray}
where $h^{(1)}$ the one-loop prepotential\footnote{The one loop
${\cal N} = 2$ four dimensional vector multiplet prepotential $h^{(1)}$ was calculated as an ansatz solution to a
differential equation involving one loop corrections to gauge coupling constant
in \cite{hamou}. However, its exact general form for any four
dimensional compactification of the heterotic string was calculated in
\cite{xri}.
Higher derivatives of $h^{(1)}$ were also calculated in \cite{lustis}.}, 
$G_i$ the orders of 
the subgroup G which leaves the i-complex plane unrotated 
and we have included the
one-loop Green-Schwarz term \cite{nistie}.
Note however, that in (\ref{duale1}) we have neglected the contribution
from the second term of (\ref{gauge1}) as it is not needed in our
present study.
Its contribution will be examined elsewhere.

The unknown dilaton dynamics, parametrized by ${\cal K}(S)$
in (\ref{duale1}), involves the contribution from the
tree level dilaton term and the non-perturbative
dilatonic contributions of (\ref{gauge1}).

Another attempt to study the classes of superpotentials of
eqn.'s 
(\ref{megas0}-\ref{megas00}) was made in \cite{homou}.
In \cite{homou} the behaviour of this class of superpotentials
was examined in the context of S-duality of ${\cal N} = 1$ heterotic string
effective action. The authors used (\ref{megas00}), when $r=-1$,
by imposing in addition, what they called,
"validity" of weak coupling perturbation theory. The latter condition
means that the superpotentials are regular anywhere except the
$S \rightarrow \infty$ limit where they can have singularities.
That enforces
the condition that the allowed forms of $W(S)$ may be in
the form\footnote{we express only the S-part of the superpotential}
\begin{equation}
W(S) =\left(\frac{1}{\eta(S)^2}\right) \left(j^{n/3}(z)(j-1728)^{m/2}(S)\right)
\left( \frac{P_1(j)}{P_2(j)}\right),
\label{megas001}
\end{equation}
or equivalently
\begin{eqnarray}
W(S)& =& \left(\frac{1}{\eta(S)^2}\right) \left( \Omega^{\prime}(S)
\right)\left( \frac{P_1(j)}{P_2(j)}\right),\\
\Omega^{\prime}(S) &= &\left(j^{n/3}(S)(z)(j(S)-1728)^{m/2}\right),
\label{megas002}
\end{eqnarray}
where m, n positive integers and $deg P_2 > deg P_1 + \frac{1}{3} n +
\frac{1}{2}m + \frac{1}{12}$, and $P_1$, $P_2$ are polynomials in j.

The class of superpotentials (\ref{megas002}) have negative
cosmological constant at its minimum $S = 1$. Moreover,
the behaviour of (\ref{megas001}), since W is a section on a flat holomorphic
line bundle over the moduli space of $SO(2)\setminus SL(2,R)/SL(2,Z)$, fixes
from the behaviour of $\Omega(S)$ under the dilatational transformation $S \rightarrow S + 1$ near
$ S \sim i\infty$,
\begin{equation}
n\;mod\;3,\;\;\;m\;mod\;2 .
\label{megas003}
\end{equation}
That means that $\Omega^{\prime}$ is constrained to be
\begin{equation}
\Omega^{\prime}_{cons}(S) = [\left(j^{n/3}(S)(j(S)-1728)^{m/2}\right)]|_{n\;mod\;3,\;m\;mod\;2}
\label{megas004}
\end{equation}

Our first reaction by looking at the factor $\Omega^{\prime}$, of (\ref{megas002}) which
is a part of S-duality invariant superpotentials and $\Omega$ of
(\ref{megas00}), that describes part of a T-duality invariant
superpotential, may be that they are quite different
as the range of parameters in the exponentials are different.
Ideally, we would expect that given the 
$\Omega^{\prime}$, $\Omega$ to
be quite the same since they have both the same modular weight 
zero. In addition, because (\ref{duale1}) should transform as a line bundle under
target space duality translational transformations at infinity, we should
expect the constraints (\ref{megas003}) to be apriori incorporated
even on the T-duality superpotentials (\ref{duale1}).\newline
In the next section, we will suggest that this is the case,
$\Omega \rightarrow \Omega^{\prime}_{cons}$, and the 
constaints
(\ref{megas003}) may be incorporated apriori in the superpotentials of
(\ref{duale1}) or (\ref{duale2}).

We recall \cite{cve} one more result, in the context of T-duality
transforming superpotentials
namely that it is possible to construct superpotentials $W^j(T, S)$
that break supersymmetry
and have vanishing cosmological constant at generic points in the
moduli space, when the auxiliary field
$h_s$ is not zero,                          
in the form
\begin{equation}
W^j(T, S) = \frac{{\cal K}(S)}{\eta^6(T)}\left( j(T)^3 +
l\;j(T)^2 + m\;j(T) + n \right),
\label{awa1}
\end{equation}
where ${\cal K}(S)$ decsribes the unknown dilaton dynamics and
$l$, $m$, $n$ constants determined by the vacuum dynamics.
The superpotentials (\ref{awa1}) have in addition minima with unbroken
supersymmetry and negative cosmological constant at the self-dual points
$T=1$, $h^T=0$.
In the following section we write down a more general class of superpotentials
that incorporates parts of approaches of \cite{cve, iba2, homou} as well
connecting the factor $\Omega^{\prime}_{cons}$ to the class of superpotentials
(\ref{awa1}).

\section{\bf New constructions of non-perturbative superpotentials}

\subsection{{\em SL(2,Z)'s modular group transforming W's}}

The method of using (\ref{megas0}) as a starting point 
to construct superpotential W's that parametrize the unknown dynamics
works
only if the modular weight r is not zero. In addition 
the
superpotential (\ref{duale1}) misses the constraint (\ref{megas003}). 
Here, we follow a different approach such that (\ref{megas003}) is included
by construction in the superpotential.  First we demand that
the behaviour of the modular weight factor\footnote{in real terms a
modified version of it} $\Omega$ factorizes from
the overall\footnote{Note that the overall
factor $\eta^{2r}$ is associated with the one loop string threshold
corrections to the gauge coupling constants \cite{kokorakos}.}
factor $\eta^{2r}$. That means that it is impossible
to describe the superpotential in terms of its form ({\ref{megas00}). 
That happens
because the construction of (\ref{megas00}) was based on a mathematical
theorem of modular forms that interconnects the factor $\eta^{2r}$
to the other factors in front involving $G_4$, $G_6$.
Our objective in this paper is to find a way of producing the
class of superpotentials of (\ref{megas004}) in such a way such
that the constraints of the monodromy behaviour of W at infinity
are included apriori.

 Our construction proceeds by writing an analogous expression
to (\ref{megas0}).
What we want to construct is the most general modular factor which
transforms with modular weight zero under SL(2,Z) modular transformations.
After we have done that we can use this factor to discuss how we can
construct superpotentials with or without singularities in the upper half
complex plane.

The clear difference in our constructions, against the approaches
of \cite{cve, iba2, homou}, is
that do not treat the
$\eta$-invariant as a fundamental quantity but
rather the cusp forms $\triangle(z)$ of the modular group $SL(2,Z)$ instead.
For this reason we write down the most general weight zero factor as
\begin{equation}
{\tilde \Sigma}(\Phi) = \frac{ E_4^i(\Phi) E_6^j(\Phi)}{\triangle(\Phi)^k},
\label{megas1}
\end{equation}
where i, j, k arbitrary integers.
In order that the modular factor ${\tilde \Sigma}(\Phi)$ to
transform
with modular weight zero the following condition must hold, 
\begin{equation}
4 i + 6 j = 12 k,
\label{megas11}
\end{equation}
with solution 
\begin{equation}
i = 3 a,\;\;\;\;j = 2 b,\;\;\;a, b\;\in\;Z
\label{megas2}
\end{equation}
Applying the constraint (\ref{megas2}) to eqn. (\ref{megas1}) we get
\begin{equation}
{\tilde \Sigma}(\Phi) = \frac{E_4^{3a}(\Phi)}{\triangle(\Phi)^a} \frac{E_6^{2b}(\Phi)}{
\triangle(\Phi)^b}
\label{megas3}
\end{equation}
or equivalently
\begin{equation}
{\tilde \Sigma}(\Phi) = \left(\frac{E_4^{3}(\Phi)}{\triangle(\Phi)}\right)^a
\left(\frac{E_6^{2}(\Phi)}{
\triangle(\Phi)} \right)^b =  j^a \cdot (j-1728)^b ,
\label{megas}
\end{equation}
where $a$, $b$ integers and j the SL(2,Z) modular function.

The non-perturbative superpotential, that includes the dilaton can now
be written into the form
\begin{equation}
W^{new}(T, S) = \frac{{\cal K}(S)\; {\tilde \Sigma}(T)}{\eta^6(T)}{\cal P}(j(T)),\;\;\;a,b\;\in\;Z,
\label{nonpert}
\end{equation}
where ${\cal P}(j(T))$ an arbitrary polynomial of the absolute modular invariant
j.

Note that the factor ${\tilde \Sigma}(T)$ that has modular
weight zero parametrizes the unknown T-modulus dynamics and incorporates
apriori by construction the constraints (\ref{megas003}).
However, there is an extra degree of freedom in the way that we could define a 
candidate T-dual superpotential since the following superpotentials have the
same modular weight, and are both allowed
\begin{equation}
W(T) =\frac{{\tilde \Sigma}(T)}{\eta^6(T)}{\cal P}(j),
\label{proto}
\end{equation}
\begin{equation}
W(T) = \frac{1}{\eta^6(T)}\frac{1}{{\tilde \Sigma}(T)}{\cal P}(j).
\label{deytero} 
\end{equation}
However, this is a novel feature
 of our construction since the candidate superpotentials (\ref{deytero})
solve the decompactification problem of the potential
(\ref{proto})(which tends to $\infty$) since they are finite at this
limit
and go to zero.

One more observation should be added here. Defining the candidate T-dual
superpotentials
into the forms (\ref{proto}), (\ref{deytero}) is equivalent  
to demanding that do not allow or do allow poles in the upper half plane
respectively. 
The associated scalar potentials are such when $T \rightarrow \infty$,
$V \rightarrow \infty, 0$ respectively. 
That happens because modular functions which are allowed
to have poles in upper half plane are exactly rational functions in j (quotients
of polynomials in j) whereas modular functions which are not allowed to have such
poles
are polynomials in j.

Let us now discuss if there is anyway that string vacua
described by the
superpotential ({\ref{megas})
are connected in anyway to the class of solutions
(\ref{awa1}) of \cite{cve}.
The answer ot this question comes from the theory of modular forms.
The expression (\ref{awa1}) is a special basis of expressing a
modular zero form in terms of a polynomial in j(z) instead of using
the more obvious basis $j(z)^n$, {$n= 0,1,2,\cdots$ in the space of
polynomials of $j(z)$.
{\em That means that the expressions of eqn.'s (\ref{megas}) and (\ref{awa1})
represent expressions for superpotentials
expressed in different basis for modular forms.}
The question as to whether string theory can fix exactly\footnote{a lower
limit on $(a, b)$ may be provided in section 4.3} the parameters
$a$, $b$ in
 (\ref{megas})
has to be decided when a non-perturbative calculative framework
in the heterotic string context is found.
The vacuum structure of the superpotentials (\ref{proto}) may be examined
by the study of the effective scalar potential V,
\begin{equation}
V = |h_s|^2 G_{S{\bar S}}^{-1} + |h_T|^2 G_{T {\bar T}}^{-1} - 3 exp(G)=
\label{dinamiko}
\end{equation}
\begin{equation}
=\frac{1}{S_R T_R^3 |\eta(T)|^{12}}\left(|S_R {\cal K}_S - {\cal K}|^2
|{\tilde \Sigma}|^2 + \frac{T_R^2}{3}|{\tilde \Sigma}_T + \frac{3}{2 \pi}
{\tilde \Sigma} {\tilde G}_2|^2 
|{\cal K}|^2 -3 |{\cal K}|^2 |{\tilde \Sigma}|^2  \right),
\label{potencial}
\end{equation}
where $h_i = exp(G/2)  $ the auxiliary field for the i-modulus.
We will not attempt to perform an extensive numerical analysis of the above
potential as this will be left for future work. We can however borrow some of the results
of \cite{cve} which are inside our range of parameters and make some comments.
In the case ${\cal P}=1$, for $(a, b) = (0, 0)$($(m, n)= (0, 0)$), that
is the case of \cite{iba1}, the minimum is at $T_{min} \approx 1.2$ while for
$(a, b)= (0,1)$ ($(m, n)=(0, 3)$), the minimum appears is at the self-dual point
$T_{min} = 1$ with the potential being negative and the other
minimum is at the
self-dual point $T= \rho = e^{i \pi/6}$ with zero potential.

Notice that the previously mentioned values have been calculated under the
condition
$S_R  {\cal K}_S - {\cal K} = 0$, $S_R = (S + {\bar S})$, which makes the
minima
to occur for weak coupling, at large $S_R$.

Note the novel feature of (\ref{deytero}) that when used to calculate
the scalar potential makes it not to diverge and tend to zero when T goes to
infinity.

\subsection{\em {\cal N} = 1 Strong-weak coupling $SL(2,Z)_S$ superpotentials}

At this point we may make some comments related to how we could
construct the most general ${\cal N} = 1$ four dimensional superpotentials that transform 
under the S-duality transformations (\ref{dual3}). Our approach generalizes
the constructions of classes of superpotentials in \cite{iba2, homou}. 
Because of its construction the modular weight zero factor
${\tilde \Sigma}(S;a, b \in Z)$ has arbitrariness as to whether
a, b are positive or negative integers or addressed in a different form
as to whether our theory may have \cite{iba2} or not singularities in the upper
half-plane or at infinity \cite{homou}.
We can choose for convenience the effective action of a possible ${\cal N} = 1$ S-dual
heterotic string action. 

In this case we might as well have as candidates
dilaton dependent superpotentials, the following\footnote{$Z^{\dagger}$ the
space of positive integers}
\begin{equation}
W(S)_{dual}^{(1)} = \frac{1}{\eta^2(S)}{\tilde \Sigma}(S) P(j(S)),\;a, b\;\in\;Z^{\dagger},
\label{der1}
\end{equation}
or
\begin{equation}
W(S)_{dual}^{(2)} = \frac{1}{\eta^2(S)}\frac{1}{{\tilde \Sigma}(S)} P(j(S)),\;a, b\;\in\;Z^{\dagger}.
\label{der2}
\end{equation}
Both (\ref{der1}), (\ref{der2}) can be classified
according as to whether we demand our superpotentials
may have or not poles in the upper half plane, equivalently
when the potential goes to zero or diverges when $S \rightarrow \infty$
respectively.  In both cases the potential is finite for finite values of S. 
In the case of (\ref{der1}) there are no poles in the upper half-plane
and the potential diverges when $S \rightarrow \infty$, while in
(\ref{der2}) the potential goes to zero when $S \rightarrow \infty$.
Thus the classes
of superpotentials (\ref{der2}) resolve naturally, the problem of
infinite potential of (\ref{der1}), when $S \rightarrow \infty$.

The superpotential (\ref{der1}) is different from the classes of
superpotentials of \cite{iba2}, where the constraint (\ref{megas003})
was not included,
as the constraint (\ref{megas003}) in our case is 
included by construction.\newline
Let us make some comments at this point regarding the
classes of superpotentials
(\ref{der2}). At first look (\ref{der2}) appears to be equivalent to the
classes of S-dual superpotentials (\ref{megas002}) of \cite{homou}.
However, (\ref{der2}) is more general. That happens partially because
the constraints (\ref{megas003}) are included by construction in its
form
and secondly because the existence of singularities
in the upper half plane at the, large dilaton limit, weak coupling region,
is enforced naturally
by construction, e.g when ${\cal P}(j(S))=1$.

\section{\bf Gaugino condensation and supersymmetry breaking}

The plan of this section is as follows. In subsect. 4.1 we reexamine
the gaugino condensation approach without the use of matter fields in the effective
lagrangian approach.  In subsect. 4.2 we reexamine gaugino condensation
in the presence of matter fields, finding all allowed forms
of truncated superpotentials.
In subsect. 4.3 we examine the stability of truncated superpotentials
of subsect. 4.1, obtaining numerical constraints on the parameters $a$, $b$ 
that are 
involved in the construction of the modular factor $\tilde \Sigma$ of
(\ref{megas}).

\subsection{\em Without matter field condensates}

We reexamine in this section earlier results \cite{fmtve,tay} on gaugino
condensation using the effective lagrangian approach
in effective supergravity theories from superstrings.
In particular we will examine the way that the superpotential of the
gaugino dynamics may
be modified when the source of non-perturbative fields of the T-field
is not specified, and defined via (\ref{megas}).
                          
We suppose that the gauge group of the ${\cal N} = 1$ supersymmetric 
superstring vacuum is a product
of non-abelian gauge factors $G_a$, $a=1,\dots,p$ and the gauge group is
given by $G = \oplus \Pi_a G_a$. 
The effective local lagrangian that incorporates p-gaugino condensates  
is then given by 
\begin{equation}
{\cal L}= - [\frac{1}{2} e^{-\frac{K}{3}} S_o {\bar S}_o ]_D
+ [S_o^3 w]_F + (f_{ab} W^a W^b)_F,
\label{ers1} 
\end{equation}
where K is the K\"ahler potential, W is the superpotential
and $f_{ab}$ the gauge kinetic function.
 The effective action
for the gaugino composite is described by defining chiral composite
superfields U, Y such that
\begin{equation} 
Y_n^3 = \frac{(\delta_{ab} W^a_{\alpha} {\epsilon}^{\alpha \beta}
W^b_{\beta})_n }{S_o^3}=\frac{U}{S_o^3}, 
\label{ers2}
\end{equation}
where the scalar components of $Y_n$'s are associated with the gaugino
condensate
$(\lambda {\bar \lambda})_n$ and the $SL(2,Z)_T$ modular weight of $Y_n$
is -1. 
The choise of the chiral compensator $S_o$ is  
such that it determines the normalization of the gravitational action
${\cal L}_{grav} = (16 \pi^2)^{-1} M_p^2 R$ with $[e^{-K/3}S_o {\bar
S}_o]_{\theta = {\bar \theta} = 0} = 1$.
The K\"ahler potential is given by
\begin{equation}
K   = - 3 \log\left( (T+ {\bar T})(S + {\bar S})^{1/3} - \sum_{n=1}^p
|Y_n|^2
\right),
\label{ers3}
\end {equation}
while the superpotential \cite{tay} is given by
\begin{equation}
w = \frac{1}{32 \pi^2} \sum_{n=1}^p c_n Y_n^3 \log[Y_n^3 \Psi_n(S) H_n^{(o)}(T)].
\label{erss4}
\end{equation}
In (\ref{erss4}) we choose to fix the value of the dilaton by using
more than one condensates
That
makes sure that a minimum at weak coupling could be found. 
 
Because of anomalous Ward identities, under the $SL(2,Z)_T$ modular
transformations (\ref{target}),
\begin{equation}
Y_n \rightarrow \frac{1}{iCT+ D} Y_n,\;\;H_n^{(o)} \rightarrow (iCT+ D)^3 H_n^{(o)},
\label{asdes5}
\end{equation}
while
\begin{equation}
\Psi_n (S) = e^{\frac{32 \pi^2 k_n S}{c_n}}.
\label{asds6}
\end{equation}
The use of $H_n^{(o)}(T)$  is such that it compensates for the lack
of modular invariance in the logarithmic term of
(\ref{erss4}).
The function $H_n^{(o)}(T)$ were given a first estimate in \cite{fmtve} as
\begin{equation}
H_n^{(o)}(T) \propto \eta^6(T)\;,
\label{asssad7}
\end{equation}
by demanding absence of singularities in the effective action in the
upper half-plane for complex T.  
We can now revise the estimate of $H_n^{(o)}$ in \cite{fmtve} by listing
all possibilities allowed depending on what kind of singularities we
admit to appear in the upper
half plane in the truncated\footnote{obtained by
integrating out the composite gaugino superfield $Y_n$.}
superpotential $W^{tr}$.\newline
We distinguish two cases:
\begin{equation}
H^{(o)} \rightarrow H^{(1)}_n(T) = \eta(T)^6 \frac{1}{j^a(T) (j(T)-1728)^b};\;\;a,b\;\in\;Z^{\dagger},
\label{asd8}
\end{equation}
which causes $W^{tr}$ to be regular in the
upper half plane,
and 
\begin{equation}
H^{(o)} \rightarrow H^{(2)}_n(T) = \eta(T)^6 {j^a(T) (j(T)-1728)^b};\;\;a,b\;\in\;Z^{\dagger},
\label{asd91}
\end{equation}
which allows  $W^{tr}$ to have poles in the upper half plane.
The modular invariant scalar potential of the theory is given by
\begin{equation}
V(S,T,Z_n) = \{32 \pi^2(1- \sum_{n=1}^p|Z_n|^2)^{-2}\}
\cdot (V_Z + V_S + V_T -3|\sum_{n=1}^pc_n Z_n^3|^2), 
\label{gaupo}
\end{equation}
where
\begin{equation}
Z_n = (S+{\bar S})^{-1/6}(T + {\bar T})^{-1/2}Y_n
\label{erote1}
\end{equation}
\begin{eqnarray}
V_Z =3 \sum_{n=1}^p c^2 |Z_n|^4 \cdot |1+\log\{(S+{\bar S})^{1/2}(T +
{\bar T})^{3/2}f_n H_n Z_n^3\}|^2\;,\\
V_S =|\sum_{n=1}^p c_n
Z_n^3\{1+ (S+{\bar S})\frac{f_{nS}}{f_n} \}|^2,\;\;
V_T =\frac{1}{3}|\sum_{n=1}^p c_nZ_n^3\{3 + (T + {\bar T})\frac{dH_{n}/dT}{H_n}\}|^2 .
\label{mega1}
\end{eqnarray}
For a minimum of the potential in the weak coupling region to exist, that is
found to be a 
minimum of a zero energy, 
this have to be determined as a solution
to the equations
\begin{equation}
\frac{\partial W}{\partial Y_n} = 0,\;\;\;\frac{\partial W}{\partial T}=0,\;\;\;
\frac{\partial W}{\partial S} = 0.
\label{asdas56}
\end{equation}
The first equation of (\ref{asdas56}) gives
\begin{equation}
\log\{Y_n^3 f_n(S) H_n(T)\}=-1 \Longrightarrow Y_n^3 =e^{-1}e^{-\frac{32\pi^2 k_n S}{c_n}}
\frac{1}{H_n^i(T)},\;i = 1,2.
\label{asdas89}
\end{equation}
That makes the truncated superpotential to take the form
\begin{equation}
W^{tr}= -\frac{e^{-1}}{32 \pi^2}K(S) h(T),
\label{asdsa21}
\end{equation}
where
\begin{equation}
K(S)=\sum_{n=1}^p c_n e^{-32 \pi^2 \frac{k_n S}{c_n}},
\end{equation}
\begin{equation}
h(T) =\frac{1}{H_n^i(T)},\;i = 1, 2. 
\label{asdsa22}
\end{equation}
Making use of the $H_n$'s in eqn.'s (\ref{asd8}, \ref{asd91})
we derive the allowed truncated superpotential forms when the
source of non-perturbative effects is not specified respectively as
\begin{equation}
W^{tr\;(1)} =  -\frac{e^{-1}}{32 \pi^2}K(S)\frac{j^a(T)
(j(T)-1728)^b}{\eta^6},\; a,b\;\in\;Z^{\dagger},
\label{sdas24}
\end{equation}
or
\begin{equation}
W^{tr\;(2)} =  -\frac{e^{-1}}{32 \pi^2}K(S)\frac{1}{\eta(T)^6} \frac{1}{j^a(T) (j(T)-1728)^b},\;a,b\;\in\;Z^{\dagger}.
\label{sdas25}
\end{equation}
The superpotentials (\ref{sdas24}), (\ref{sdas25})        
avoid and have singularities in the fundamental domain respectively.
The potential corresponding to the superpotentials of
eqn.'s (\ref{sdas24}, \ref{sdas25}) does have a stable minimum with
respect to the S-field. In particular the dilaton can be stabilized at the weakly coupled
regime as we are using more than one condensates in the way suggested in
the racetrack models of \cite{kra,la}.  
The value of the T-moduli may be found by minimizing the potential
arising
from substituting the value of the auxiliary field Y  
in (\ref{gaupo}).  Equivalently, we can use 
the values of the truncated superpotentials 
(\ref{sdas24}, \ref{sdas25}) in (\ref{dinamiko}) and minimize
with respect to the dilaton and T moduli.

\subsection{\em With matter field condensates}

Consider a heterotic string compactification vacuum such that
Wilson line background fields are associated to chiral matter fields $A$.
The latter
break the hidden sector gauge group to an $SU(N)$
hidden sector representing SQCD with M massive flavours\footnote{Q
represent the matter fields}
$Q \oplus {\bar Q}$, in the representations $N \oplus {\bar N}$, $N$ being the number of
flavours.

Recall again that the effective action in
the usual supergravity was given in eqn. (\ref{ers1}).
In the presence of matter fields the effective superpotential
as dictated by Ward identities and modular invariance has been written
in \cite{luta} as
\begin{equation}
W^{matter} = \frac{1}{32 {\pi}^2} Y^3 \log\{ e^{32 {\pi^2} S}
[ (c \eta(T)]^{6})^{N - M/3} Y^{3N -2M} det \Pi\} - tr A \Pi,
\label{sdas26}
\end{equation}
Here, c is an unknown constant, $A$ chiral matter fields associated with the
Wilson line background fields and $\Pi_i^j = Q_i {\bar Q}^j$, $j=1,\dots,M$ ,
represent the matter bound states.
We would like now, given the constructions of $W^{non-pert}$ originating from
eqn. (\ref{megas3}), to see if we can modify somehow eqn. (\ref{sdas26})
to incorporate the unknown non-perturbative T-duality dynamics in all
possible forms.
The possible modifications of (\ref{sdas26}) read
\begin{eqnarray*}
W^{matter(1)} = \frac{1}{32 {\pi}^2} Y^3 \log\{ e^{32 {\pi^2} S}
[ (c \eta(T)]^{6})^{N - M/3} \left(\frac{1}{j^a(T) (j(T)-1728)^b}
\right)^{N - M/3}
 Y^{3N -2M} \\
 \times det \Pi\}
 - tr A \Pi,\;a,b\;\in\;Z^{\dagger},
\end{eqnarray*}
\begin{equation}
\label{sdas26a}
\end{equation}
or alternatively
\begin{eqnarray*}
W^{matter(2)} = \frac{1}{32 {\pi}^2} Y^3 \log\{ e^{32 {\pi^2} S}
[ (c \eta(T)]^{6})^{N - M/3} \left({j^a(T) (j(T)-1728)^b}
\right)^{N - M/3}
 Y^{3N -2M} \\
 \times det \Pi\}
 - tr A \Pi,\;a,b\;\in\;Z^{\dagger}.
\end{eqnarray*}
\begin{equation}
\label{sdas27}
\end{equation}
Because at the weak coupling limit, $ReS \rightarrow \infty$, gravity
decouples and the string model behaves like globally supersymmetric QCD,
global supersymmetry is not broken and the minimum of the potential
is found by the conditions
\begin{equation}
\frac{\partial W}{\partial Y}= \frac{\partial W}{\partial \Pi}= 0.
\label{aux}
\end{equation}
Solving eqn.'s (\ref{aux}) results in
\begin{eqnarray}
\frac{1}{32 \pi^2} Y_{(1)}^3 &=& 32 \pi^2 e^{M/N -1}[c \eta]^{2\frac{M}{N} -6}
  \{j^a (j-1728)^b\}^{1- \frac{M}{3N}}] [det A]^{1/N}exp{(-32 \pi^2 S/N)},\nonumber\\
  \Pi &=& \frac{1}{32 \pi^2} Y_{(1)}^3 A^{-1},\;a,b,\;\in\;Z^{\dagger},
\label{auxi1}
\end{eqnarray}
or
\begin{eqnarray}
\frac{1}{32 \pi^2} Y_{(2)}^3 &=& 32 \pi^2 e^{M/N -1}[c \eta]^{2\frac{M}{N} -6}
  \{j^a (j-1728)^b\}^{1- \frac{M}{3N}}] [det A]^{1/N}exp{(-32 \pi^2 S/N)},\nonumber\\
  \Pi &=& \frac{1}{32 \pi^2} Y_{(2)}^3 A^{-1},\;a,b,\in\;Z^{\dagger},
\label{auxi2}
\end{eqnarray}
respectively.
We can now
eliminate the auxiliary composite fields $Y$, $\Pi$ in eqn's (\ref{sdas26a}),
(\ref{sdas27}),
by substituting their values from (\ref{auxi1}), (\ref{auxi2})
respectively and derive
the truncated superpotential $W^{tr}_{matter}$, which depends only on the
S, T and A fields.
The truncated superpotential reads for the two cases considered
\begin{eqnarray}
W_{trunc}^{(1)} &=& {\hat \Omega}(S) {\cal K}(T)[det A]^{1/N}\\
{\hat \Omega}(S) &=& -N exp(-32 \pi^2 S/N) \\
{\cal K}(T) &=& (32 \pi^2 e)^{M/N-1} [c \eta(T)]^{2M/N-6} {\cal P}^{(k)}(j(T)),\;k=1,2 
\label{tranc}
\end{eqnarray}
where
\begin{equation}
{\cal P}^{(1)} = j^a(j-1728)^b,\;a,b,\;\in\;Z^{\dagger},
\label{ena}
\end{equation}
\begin{equation}
{\cal P}^{(2)} = \frac{1}{j^a(j-1728)^b},\;a,b,\;\in,Z^{\dagger}.
\label{dio}
\end{equation}
The superpotentials associated with (\ref{ena}), (\ref{dio}) represent candidate
solutions
for non-perturbative superpotentials in the presence of the "Wilson
lines" A.

\subsection{\em Stability of gaugino condensate in the truncated 
 formalism}

To test whether or not the new constructions $W^{non-pert}$
of eqn. (\ref{sdas24}) 
might be phenomelogically preferred over the eqn. (\ref{sdas25})
form\footnote{something similar may be shown in the two cases in
(\ref{tranc}), of 
$k=1$ over the $k=2$ case, when matter field condensates
are included,},
we will look at the 
stabilization conditions for the gaugino condensate. For convenience
we look at the case without matter field condensates of subsection 4.1.
For convenience we choose the K\"ahler 
potential in the form
\begin{eqnarray}
K_{pert}= -\log(S + {\bar S}) - 3\log((T + {\bar T}),\nonumber\\
K = K_{pert} - 3 \log(1 - \frac{9}{\psi} e^{\frac{K_{pert}}{3}}(Y {\bar
Y})^{1/3}),
\label{syme}
\end{eqnarray}
where\footnote{Note that 
the effective action in this form was singled out in \cite{bureo}.
By setting $\psi = 9$ we recover the normalization of \cite{tay}.
}
$\psi$ a constant. 
The effective potential of the theory is given by
\begin{equation}
V = \frac{b}{6}\frac{|\lambda|^4}{(1-|{\tilde z}|^2)^2)^2}\{3|1 + 
\ln(c {\cal B}(S) H^k(T)e^{-K/2}{\tilde z}^3)|^2 + {\cal E} |{\tilde z}|^2\},
\label{defi}
\end{equation} 
where $\cal B(S)$ describes the gaugino condensation dynamics of the dilaton,
\begin{equation}
{\tilde z} = Y exp{(K_{pert}/2)},  
\label{defi1}
\end{equation}
$\tilde z$ the modified Z variable$(Z={\tilde z}$) of (\ref{erote1}),
$H^k(T)$ is taken from from (\ref{asd8}) and (\ref{asd91})
and 
\begin{equation}
{\cal E } = (S + {\bar S})^2 \left|- \frac{1}{(S + {\bar S})} +
\frac{{\cal B}^{\prime}}{\cal B} \right|^2
+ \frac{1}{3}\left|3 + (T + {\bar T})\frac{H^{\prime}(T)}{H(T)}\right|^2-3,
\label{sdaqw12}
\end{equation}
From the solution of eqn.'s (\ref{asdas56}), that describe the minimum of the potential at the
weak coupling limit
we can calculate the value of the 
gaugino condensate Y as
\begin{equation}
|<\lambda \lambda>| =|<Y>| = e^{-1/3}\frac{(c {\cal B}(S) H(T))^{-1/3}}{(T + {\bar T})^{1/2}},\;\;
{\tilde z}_{min}=  e^{-1/3}\frac{(c {\cal B}(S) H(T))^{-1/3}}{(S+ {\bar
S})^{1/6} (T + {\bar T})^{1/2}}.
\label{defi200}
\end{equation}
The stationary point of the potential (\ref{defi}), a zero energy 
minimum, is reached when ${\cal E} \rightarrow 0$ and in addition
$S_R {\cal B}^{\prime} - {\cal B}=0$.
The last condition is necessary for the potential to achieve
a minimum at the weak coupling region \cite{iba1}. In this case, 
in the absence of a well defined dilaton dynamics, a minimum at weak coupling
could be easily achieved for the superpotential class of \cite{dine}, 
${\cal B}(S)= c^{\prime} + h^{\prime}_{\tilde a} e^{\frac{3S}{2b_{\tilde a}}}$, where
$\tilde a$ counts the number of condensates, when the parameter
$c^{\prime}$ is very small. 
 Exactly as in \cite{fmtve}
the minimum of the potential is found when the phase of the condensate
is aligned in such a way that the quantity $(c {\cal B}(S)
H^k(T)e^{-K/2}{\tilde z}^3)$ is real and positive.  

Notice that we are interested in the minimum of the potential at weak coupling
because phenomenology requires unification of the gauge couplings of the
standard model at a unification scale of $10^{16}$ GeV. In
particular(see for example \cite{de})
this requires $S_R \approx 2$.  
The behaviour of the potential (\ref{defi}) as to whether it reaches its
stationary point   
 depends crucially on the behaviour of the parameter 
${\cal I} \equiv {\cal E}|{\tilde z}|^2 $. 

Notice that we make here an important point. 
We require that in order for the potential to reach the zero energy minimum 
the parameter ${\cal I} $ has to vary smoothly at a general
point
of the moduli space. Since the dilaton dynamics is in general unknown,
we assume that $S_R {\cal B}^{\prime} - {\cal B}$ goes to zero smoothly.
Separating the T-modulus dependence we find that
\begin{equation}
{\cal I} \sim (T + {\bar T})\; \frac{|H^{\prime}(T)|^2}{|H(T)|^2}\;H(T)^{-2/3}.\;
\label{sdfg123}
\end{equation}
We rewrite the value of H from (\ref{asd8}), (\ref{asd91}) into the
form
\begin{equation}
H(T)= \eta^6(T)\frac{1}{j^a(j-1728)^b},\;\;\;a,b\;\in\,Z.
\label{nb1}
\end{equation}
The only points in the moduli space that $\cal I$ may have a problem is
exactly the self-dual, duality invariant points of the 
potential (\ref{defi200}) $T = 1,\;\rho$.
To illustrate the nature of singularities in the moduli space we notice
that j has a zero of order 3 at $T=\rho$ and (j-1728) has a zero of order
2 at $T = 1$.
That means 
\begin{equation}
(j-1728) \sim (T-1)^2\;when\;T\;\rightarrow\;1,\;j\;\sim\;(T-\rho)^3\;when\;T\;\rightarrow\;\rho.
\label{vb12}
\end{equation} 
In addition,
\begin{equation}
\frac{H^{\prime}}{H} \stackrel{T \rightarrow \rho}{\rightarrow} (T-\rho)^{-2},\;\;\;
\frac{H^{\prime}}{H} \stackrel{T \rightarrow 1}{\rightarrow} (T-1)^{-2}\;.
\label{etas}
\end{equation}
Before examining the effect of (\ref{nb1}) to (\ref{sdfg123}) let us produce the 
points where the potential corresponding to the class of superpotentials
(\ref{megas00})
blows up.  
As $T \rightarrow \rho$ we get \cite{dent} that (\ref{sdaqw12}) breaks 
down for $m  \leq 2$.
Notice that if we perform a similar analysis for the superpotentials
(\ref{megas00}) as $T \rightarrow 1$ we get that
\begin{equation}
{\cal I} \sim (T-1)^{-2 + 2m/3},
\label{jkl12}
\end{equation}
That means that (\ref{defi200}) breaks down\footnote{goes to infinity}, as $T \rightarrow 1$
for the range of parameters $n \leq 2$.
Summarizing
\begin{equation}
V|_{j^{n/3} (J-1728)^{m/2}} \stackrel{T \rightarrow \rho}{\rightarrow}\infty,\;\;n \leq 2,\\
\;V|_{j^{n/3} (J-1728)^{m/2}} \stackrel{T \rightarrow 1}{\rightarrow}\infty,\;\;m \leq 2.
\label{telikos}
\end{equation}

In our case, as $T \rightarrow 1$, $\cal I$ changes as
\begin{equation}
{\cal I}|_{j^a (j-1728)^b} \sim (T-1)^{4b/3 - 2}.
\label{oi12}
\end{equation}
Clearly, the values of $b$ which the potential may avoid
to develop a singularity as $T \rightarrow 1$ are
\begin{equation}
b \leq 1.
\label{zxas1}
\end{equation}
\newline
A similar analysis can be performed for (\ref{nb1}) as $T \rightarrow \rho$.
In this case
 \begin{equation}
{\cal I}|_{j^a (j-1728)^b} \sim (T-\rho)^{2a - 2},
\label{oi14}
\end{equation}
which means that $\cal I$ does not become infinite when $a$ avoids the values
\begin{equation}
a \leq 0.
\label{oi16}
\end{equation}

Note that
the upper safe value $m=3$ in (\ref{telikos}) corresponds
exactly to the value $b = 1$ of the superpotentials (\ref{nb1}). That means
that (\ref{nb1}) naturally does not make the potential to
breaks down as  $T
\rightarrow \rho$.

Thus we have derived the allowed $(a, b)$ values for the potential
to be finite\footnote{that demands the parameter
$\cal I$ to vary smoothly over the whole moduli space} at every point in the moduli
space so that the weak
coupling minimum can be reached, to be
\begin{equation}
(a ; b)=(1,2,3,...; 2, 3,...).
\label{oi18}
\end{equation}
The "allowed" values (\ref{oi18}) in the parameter space $(a, b)$
constitute a criterion for the potential to avoid singularities
in the fundamental domain.
Notice, that for those values of $(a, b)$ we should be able to 
find a minimum at weak coupling.
In addition, it is possible in this case that the minimum is 
not necessarily at 
weak coupling
because now the parameter $S_R {\cal B}^{\prime} - {\cal B}$
is allowed to varied smoothly, with no T-dependence, and it is not
necessary that it reaches zero.

It is worth noticing that for the minimum allowed values of $a$, $b$, 
$min(a, b) = (1, 2)$ an interesting possibility arises.
For those values
\begin{equation}
W(T,S, a=1, b=2)= {\cal K}(S)\frac{1}{\eta^6(T)}\;j\;(j-1728)^2 \equiv {\cal
K}(S)\frac{1}{\eta^6(T)}(j^3 - 3456 j^2 + 1728^2 j).
\label{oi20}
\end{equation}
It is clear that (\ref{oi20}) has an identical form to (\ref{awa1}), the solution
which can
have
a vanishing cosmological constant and broken supersymmetry at a generic point in  the moduli space,
for special values of the parameters l, m, n and broken supersymmetry,
\begin{equation}
W(T,S;min(a)=1, min(b)=2)  =
W^j(T,S)|_{l=-3456, m=1728^2, n=0}\;\;.
\label{oi22}
\end{equation}
Since the solutions (\ref{oi20}) represent a particular
basis for modular forms we expect (\ref{oi20}) and (\ref{awa1}) to be
equivalent.

Note that despite that fact that the gauge kinetic function develops
singularities at the self-dual points $T = 1, \rho$ signalling the
appearance of 
previously massive states becoming massless,
the potential is finite at these points.
In particular, the nature of the construction (\ref{proto}) is such
that the potential naturally avoids to blow up at $T = \rho$ while at
$T \rightarrow 1$, we enforce it to behave smoothly by restricting
the $b \geq 2$.
This is a novel feature of
our potential (\ref{proto}) since the
superpotentials (\ref{duale1}) of \cite{cve} make the potential
to become infinite when $(m, n) \leq 2$.

\section{Conclusions}

Because the degeneracy of the dilaton and the rest of the moduli
fields have to be lifted by non-perturbative effects
we studied the most general parametrizations of superpotentials such as 
the origin of non-perturbative effects is not specified in the context
of ${\cal N} = 1$ four dimensional heterotic string.

The dilaton dependence of superpotentials was examined
only in the context of a ${\cal N} = 1$ strong-weak coupling
duality equivalence \cite{iba2} of the ${\cal N} = 1$ heterotic string actions. 
In this case, we derived the most general parametrizations of
non-perturbative dynamics that do, or not, allow singularities in the fundamental
domain generalizing results by \cite{iba2, homou}.
All available S-duality constructions\cite{iba2, homou} are "included" in our
construction.

Because classes of dilaton superpotentials like (\ref{der1}), tend to
stabilize the dilaton at $S = 1$, it is worth exploiting the possibility
that our dilaton construction is generalized to subgroups of $SL(2,Z)_S$
such as those appearing in supersymmetric Yang-Mills \cite{sei, kokos1}. 
and in compactification of F-theory on $K_3$ surfaces
when the Mordell-Weyl group is not trivial \cite{aspimo}.

In another direction we found the most general parametrization of
T-moduli
dynamics, generalizing and complementing previous approach \cite{cve}.
In fact what we did, was to allow for the most general parametrization
of the corrections to the K\"ahler class T-moduli in the gauge kinetic 
function f. 
We found that the nature of the superpotential, by construction,
incorporates a novel criterion for avoiding singularities in the 
fundamental domain
or places them outside the latter, e.g expressions
(\ref{sdas24}), (\ref{sdas25}) respectively. To lowest order, in the parameters
space $(a,b)$, the finite scalar potential may correspond to vacua with vanishing
cosmological constant and broken supersymmetry, dilaton auxiliary field not zero,
found before \cite{cve}
in a different context.
Moreover, for this range of $(a, b)$ parameters it is possible that the
supersymmetry
breaking minimum can break supersymmetry when the dilaton auxiliary
field
is not zero.\newline
Anothr novel property of the superpotentials
(\ref{deytero}), for the perturbative heterotic string, 
is that the
potential goes to
to zero at the decompactification limit $T = \infty$.

Independently of what will be the exact form of the non-
perturbative superpotential for ${\cal N} = 1$ heterotic string vacua, when we will be able to calculate it
exactly, the superpotentials (\ref{megas00}),  captured the general
quantitative structure of non-perturbative
effects since they are constructed based on its modular properties.
Our results have important consequences for
supersymmetry breaking, CP violation,
and inflation \cite{cosmo} in the context of ${\cal N} = 1$ four dimensional heterotic string theories.
Such problems may require further study.

\section{Acknowledgements}
We would like to thank Don Zagier and Tom Taylor for discussions.
We would also to thank the Department of Mathematics at Michigan University,
where part of this work was done,
for its hospitality and financial support.
We are also grateful to Alexander Onassis Public Benefit Foundation
for its generous financial support of our visit at Ann-Arbor Michigan.
After this work was finished we realized that a similar
expression to (\ref{megas1}) appears in \cite{kobli}.

\end{document}